\documentclass[osajnl,twocolumn,showpacs,superscriptaddress,10pt]{revtex4-1} 
\usepackage{amsmath,amssymb,graphicx,hyperref}
\newcommand{\be}{\begin{equation}}
\newcommand{\ee}{\end{equation}}
\newcommand{\ber}{\begin{eqnarray}}
\newcommand{\eer}{\end{eqnarray}}
\newcommand{\de}{\end{equation*}}
\newcommand{\cer}{\begin{eqnarray*}}
\newcommand{\der}{\end{eqnarray*}}
\begin{document}

\title{Propagation of an arbitrary vortex pair  through an astigmatic optical system and determination of its net topological charge}

\author{Salla Gangi Reddy}\email{Corresponding author: sgreddy@prl.res.in}
\author{Shashi Prabhakar}\email{shaship@prl.res.in}
\author{Aadhi A}\email{aadhi@prl.res.in}
\author{J. Banerji}\email{jay@prl.res.in}
\author{R. P. Singh}\email{rpsingh@prl.res.in}

\affiliation{Physical Research Laboratory, Navrangpura, Ahmedabad, India - 380009}

\begin{abstract}
We embed a pair of vortices with different topological charges in a Gaussian beam and study its evolution through an astigmatic optical system, a tilted lens. The propagation dynamics is explained by a closed-form analytical expression.  Furthermore, we show that a careful examination of the intensity distribution at a predicted position past the lens can provide us with the net charge present in the beam. To the best of our knowledge, our method is the first non-interferometric technique to measure the net charge of an arbitrary vortex pair. Our theoretical results are well supported by experimental observations.
\end{abstract}


\maketitle 

\section{Introduction}
Optical vortices have drawn considerable attention in science and engineering due to their dark core and helical wave front. An optical vortex of order $l$ centered at the origin ($r = 0$) has a field distribution of the form $E(r) \exp(il\phi)$. The distribution is such that the field intensity tends to zero as $r\rightarrow0$ whereas the phase shift in one cycle around the origin is $2l\pi$ where $l$ is an integer. The azimuthal mode index $l$, also called topological charge of the vortex, has a physical meaning in that the vortex carries an orbital angular momentum (OAM) of $l\hbar$ per photon \cite{allen}. This angular momentum can be imparted to microscopic particles in order to manipulate them optically \cite{particle, particle1, grier}. In recent years, the OAM of light has also found applications in classical \cite{siddu} as well as quantum communication \cite{OAM, torner}. These applications have led to considerable interest in the generation and study of optical vortices both in free space \cite{free, spiral, cgh} and in guided media \cite{guide, guide1}.

The propagation of a pair of vortices has gained a lot of interest in research since the last decade \cite{dipole1, dipole2, dipole3, dipole4, dipole5, dipole6}. Indebetouw studied the propagation of an array of vortices through free space and showed that the relative separation between the vortices is invariant during the propagation in the case of same type (sign) of charges whereas they will attract and annihilate each other  in the case of oppositely charged vortices \cite{dipole1}. Chen and Roux have studied the annihilation of dipole vortices during their propagation. They found that the background phase function at a point where two dipoles annihilate, have a continuous potential which causes the annihilation. They have used the same background phase function to accelerate the annihilation process \cite{dipole4}. Recently, the tight focusing properties of a pair of vortices have been investigated theoretically \cite{dipole5, dipole6}. However, it dealt only with isopolar and dipole vortices of first order. Here, we present a theoretical analysis of propagation of an arbitrary vortex pair passing through an astigmatic optical system and verify the results with experiments. We suggest that the vortices being generic to all the waves \cite{berry}, the present study can be useful to acoustic \cite{thomas} and matter waves \cite{angom} also.

Since the topological charge of a vortex determines its OAM, an accurate measurement of the topological charge is an essential and important task. There are a number of methods to determine the charge of an optical vortex and its sign \cite{inter, inter1, inter2, pravin1, shashi, hick, mourka, pravin2}. For a multi-singular beam, however, there is no method to measure the net charge which determines the torque imparted by the field. We show that the intensity distribution of a multi-singular beam at a predicted position beyond a tilted lens can provide information about the net charge present in it.

\section{Theory}
Consider a pair of optical vortices embedded in a Gaussian beam, one with topological charge $\epsilon_1 m\, (\epsilon_1=\pm 1)$ located at $x_1=- x_0$, $y_1=0$ and another with topological charge $\epsilon_2 n\, (\epsilon_2=\pm 1)$ at $x_1= x_0$, $y_1=0$.  The complex field distribution of the vortex pair at the waist plane of the host Gaussian beam, with waist size $w_0$, is given by
\begin{equation}
\begin{split}
E_1(x_1,y_1)=(x_1+ x_0+ & i\epsilon_1 y_1)^{m}  (x_1- x_0+i\epsilon_2 y_1)^{n} \\
& \times\exp \left[-\left(\frac{x_1^{2}+y_1^{
2}}{{w_0^{2}}}\right)\right].
\end{split}
\end{equation}

The tilted lens is placed at a distance $z_0$ from the waist plane. The vortex passes through the lens and travels a further distance $z$.
The overall ray transfer matrix ${\bf M_{tot}}$ is given by \cite{pravin2}
\begin{equation}
{\bf M_{tot}}=\left(\begin{array}{cc} {\bf A} & {\bf B}\\-{\bf C}/f & {\bf D}\end{array}\right)
\end{equation}
where ${\bf A}$, ${\bf B}$, ${\bf C}$ and ${\bf D}$ are $2\times 2$ diagonal matrices with diagonal elements given by $a_j$, $b_j$, $c_j$ and $d_j$ respectively. Explicitly,
\begin{eqnarray}
c_1 & = & \sec\theta,\; c_2=\cos\theta,\;a_j=1-z c_j/f,\nonumber\\
d_j & = &1-z_0 c_j/f,\qquad b_j=z_0+z d_j, {j=1,2}.
\end{eqnarray}
Next, we define two column vectors ${\bf r_1}$, ${\bf r_2}$ so that their transposes are given by row vectors ${\bf r_i}^T=(x_i,y_i), {i=1,2}$.
The field $E_2(x_2,y_2)$ at a distance $z$ past the lens is given by the generalized Huygens-Fresnel integral \cite{sieg}:
\begin{equation}
E_2(x_2,y_2)= \frac{i/\lambda}{| B|^{1/2}}\int\!\!\!\int
dx_1\,dy_1\,E_1(x_1,y_1) e^{-(i\pi/\lambda){\bf \phi(r_1,r_2)}}
\end{equation}
where $| B|=| b_1b_2|$ is the determinant of ${\bf B}$ and
\begin{eqnarray}
{\bf \phi(r_1,r_2)}& = & {\bf r_1}^T {\bf B}^{-1}{\bf A}{\bf r_1}+
{\bf r_2}^T {\bf D}{\bf B}^{-1}{\bf A}{\bf r_2}-2{\bf r_1}^T
{\bf B}^{-1}{\bf r_2}\nonumber\\
& = & x_1^2 a_1/b_1+y_1^2 a_2/b_2+x_2^2 d_1/b_1+y_2^2 d_2/b_2\nonumber\\
&& \mbox{} -2(x_1x_2/b_1+y_1y_2/b_2).
\end{eqnarray}

The integration over $x_1$ and $y_1$ are carried out by writing $E_1(x_1,y_1)$ as
\begin{subequations}
\ber
E_1(x_1,y_1) & = & \lim_{\substack{t\rightarrow 0\\t'\rightarrow 0}} \left[\frac{\partial^m}{\partial t^m}\frac{\partial^n}{\partial {t'}^n}\exp\left\{f\left(t,t'\right)\right\}\right],\\
f(t,t') & = & t(x_1+x_0+  i\epsilon_1 y_1)+t'(x_1- x_0+i\epsilon_2 y_1) \nonumber\\
&& {}-\frac{x_1^{2}+y_1^{2}}{w_0^{2}}.
\eer
\end{subequations}
Using the definition of Hermite polynomial and a recurrence relation
\begin{subequations}
\ber
H_n(x) & = & \frac{\partial^n}{\partial {t}^n}\exp(2xt-t^2)\vert_{t=0}\\
\frac{d^j}{dx^j}H_n(x) & = & \frac{2^j n!}{(n-j)!}H_{n-j}(x)
\eer
\end{subequations}
we finally get
\begin{subequations}
\begin{equation}
\begin{split}
E_2(x_2,y_2)  = & \frac{kw_1w_2(i/2)^{m+n+1} \gamma^{m+n}}{(b_1b_2)^{1/2}} \\ &\,\times\exp\left[-\left(\beta_1 x_2^2+\beta_2 y_2^2\right)\right] F_{m,n}(x_2,y_2),
\end{split}
\end{equation}

\begin{equation}
\begin{split}
F_{m,n}(x_2,y_2)  =  & \sum_{j=0}^{\min(m,n)}  {m \choose j}\,{n \choose j} \Delta^j j!  \\ &\,H_{m-j}[f_1(x_2,y_2)]H_{n-j}[f_2(x_2,y_2)]
\end{split}
\end{equation}
\end{subequations}
where, $k=2\pi/\lambda$,
\begin{subequations}
\ber
\frac{1}{w_j^2} & = & \frac{1}{w_0^2}+i\frac{ka_j}{2b_j},\\
\gamma & = & (w_1^2-w_2^2)^{1/2},\\
\Delta & = & -2(w_1^2-w_2^2\epsilon_1\epsilon_2)/\gamma^2,\\
\alpha_j & = & \frac{k w_j^2}{2b_j}, \\
\beta_j & = & \left(\frac{kw_j}{2b_j}\right)^2+i\frac{k d_j}{2b_j},
\eer
\end{subequations}
and
\begin{eqnarray}
\left[\begin{array}{c}f_1(x_2,y_2)\\f_2(x_2,y_2)\end{array}\right]&=&\frac{1}{\gamma}\left[\begin{array}{c}
\alpha_1x_2+i(\epsilon_1 \alpha_2 y_2- x_0)\\
\alpha_1x_2+i(\epsilon_2 \alpha_2 y_2+ x_0)
\end{array}\right] \nonumber \\ &=& \frac{1}{\gamma}\left[\begin{array}{c}\phi_1(x_2,y_2)\\\phi_2(x_2,y_2)\end{array}\right]
\end{eqnarray}

Eqs. (8-10) form one of our main results. It generalizes previous work \cite{dipole6} on the propagation dynamics of a vortex pair through an astigmatic system in that the topological charges $m$ and $n$ need not be the same and can have arbitrary integer values.

Before proceeding further, we note that the above general result includes the following  special cases:
(1) For $m=n$, we get the propagation dynamics of (a) an isopolar vortex pair if $\epsilon_1 \epsilon_2=1$ and (b) a vortex dipole if $\epsilon_1 \epsilon_2=-1$; (2) For $n=0$, the $j$-sum reduces to the $j=0$ term only, and we get the propagation dynamics for an off-center single vortex given by
\ber
E_2(x_2,y_2)& = &  \frac{kw_1w_2(i/2)^{m+1}}{(b_1b_2)^{1/2}}\exp\left[-\left(\beta_1 x_2^2+\beta_2 y_2^2\right)\right]\nonumber\\
&&{}\times \gamma^{m} H_m[(\alpha_1 x_2+i\epsilon_1 \alpha_2 y_2-i x_0)/\gamma].
\eer
(3) Setting $x_0=0$ in the above result, one immediately recovers our previous result \cite{pravin2} for a single vortex at the origin.

The sum $F_{m,n}$ can be evaluated formally as follows. We introduce the 2-variable Hermite-Kamp\'{e} de F\'{e}riet polynomials $H_n(x,y)$ as \cite{khan}
\be
H_n(x,y)=n!\sum_{r=0}^{[n/2]} \frac{x^{n-2r} y^r}{(n-2r)!r!}
\ee
in terms of which the classical Hermite polynomials $H_n(x)$ are given by
\be
H_n(x)=H_n(2x,-1).
\ee
Next, we consider the 4-variable 2-index 1-parameter Hermite polynomials $H_{m,n}(x,z;y,w|\tau)$ defined as \cite{khan, khan3}
\be
\begin{split}
H_{m,n}(x,z;y,w|\tau)= & \sum_{s=0}^{\min(m,n)}  \tau^s s! {m \choose s}\,{n \choose s}\, \\ & H_{m-s}(x,z)H_{n-s}(y,w).
\end{split}
\ee
It is then easy to show that
\be
F_{m,n}=H_{m,n}(2f_1, -1;2f_2,-1|\Delta)
\ee
which has the following generating function
\be
\begin{split}
\exp[ & -(u^2+v^2)+2(f_1u+f_2v)+\Delta uv]\\ & =\sum_{m,n=0}^\infty \frac{u^m v^n}{m!n!} H_{m,n}(2f_1, -1;2f_2,-1|\Delta)
\end{split}
\ee
\subsection{Determination of net topological charge}
As noted earlier \cite{pravin2}, the modulations due to the Hermite polynomial become most prominent when $w_2=w_1^*$. This happens at a certain value $z=z_c$. To determine $z_c$ and {\it also} the distance $z_0$ between the waist plane and the lens, we impose the following conditions:
\be
\frac{ka_1}{2b_1}\vert_{z=z_c}=-\frac{ka_2}{2b_2}\vert_{z=z_c}=\frac{1}{w_0^2}
\ee
Solving Eqs. (17) and introducing the Rayleigh range $z_R=k w_0^2/2$,   we get
\ber
z_0 & = & z_R \left(1+\frac{2f \cos\theta}{z_R\sin^2\theta}\right)^{1/2}\nonumber\\
z_c & = & \frac{z_R (1+\cos^2\theta)+z_0 \sin^2\theta}{2(z_R/f) \cos\theta -\sin^2\theta} \label{zc}
\eer
The first equality in Eqs. (17) ensures that $w_2=w_1^*$ at $z=z_c$ (see Eq. 9a) whereas
 the last equality makes many expressions appearing in Eqs. (8-10) considerably simpler at $z=z_c$. Thus, at $z=z_c$,
\ber
\Delta & = & \left\{\begin{array}{ll}
-2 & \mbox{if $\epsilon_1 \epsilon_2=1$},\\
-2i & \mbox{if $\epsilon_1 \epsilon_2=-1$};
\end{array}
\right.\nonumber\\
\left(\begin{array}{c}w_1^2\\w_2^2\end{array}\right) & = & \frac{w_0^2}{\sqrt{2}}\left(\begin{array}{c}\exp(-i\pi/4)\\\exp(i\pi/4)\end{array}\right);\nonumber\\
\gamma & = & w_0\exp(-i\pi/4);\nonumber\\
f_1 & = & \delta_1 x_2-\epsilon_1 \delta_2 y_2 +(x_0/w_0) \exp(-i\pi/4)\nonumber\\
f_2 & = & \delta_1 x_2-\epsilon_2 \delta_2 y_2 -(x_0/w_0) \exp(-i\pi/4)
\eer
where
\be
\delta_j=\frac{kw_0}{2\sqrt{2}b_i}
\ee
\subsubsection{Vortices with topological charges of the same sign}
Suppose $\epsilon_1=\epsilon_2=1$. Then, $f_1=\theta_-+\theta_0$ and $f_2=\theta_--\theta_0$ where,
\ber
\theta_- & = & \delta_1 x_2-\delta_2 y_2\nonumber \\
\theta_0 & = & (x_0/w_0) \exp(-i\pi/4).
\eer
Note that the dependence on $x_2$ and $y_2$ is in the form $\theta_-$ only.

For a small separation between the vortices, one can expand the Hermite polynomials appearing in Eq.(8) as functions of $x_0/w_0$ by using the formula
\be
H_n(x+y)=H_n(x)+2 ny H_{n-1}(x)+O(y^2).
\ee
Substituting in (8) and using the summation rule \cite{magnus}
\be
\sum_{r=0}^{\min(m,n)}(-2)^r r! {m \choose r}\,{n \choose r}\, H_{m-r}(x)H_{n-r}(x)= H_{m+n}(x),
\ee
we get
\be
F_{m,n}=H_{m+n}(\theta_-)+2\theta_0 (m-n) H_{m+n-1}(\theta_-)+O(\theta_0^2).
\ee
For $\epsilon_1=\epsilon_2=-1$, $\theta_-$ will change to $\theta_+ =\delta_1 x_2+\delta_2 y_2$ in the above expressions.
\subsubsection{Vortices with topological charges of opposite signs}
Suppose $\epsilon_1=1$, $\epsilon_2=-1$. In this case,
\ber
f_1 & = & \theta_- +\theta_0\nonumber\\
f_2 & = & \theta_+ -\theta_0.
\eer
Note that in this case, the dependence on $x_2$ and $y_2$ is in the form $\theta_{\pm}= \delta_1 x_2 \pm \delta_2 y_2$.

For a small separation between the vortices, we can proceed as in the previous section, to get
\ber
\lefteqn{F_{m,n}=H_{m,n}(2\theta_-, -1;2\theta_+,-1|-2i)}\nonumber\\
&&{}+ 2 \theta_0 H_{m-1,n}(2\theta_-, -1;2\theta_+,-1|-2i) \nonumber\\
&&{}-2 \theta_0 H_{m,n-1}(2\theta_-, -1;2\theta_+,-1|-2i)\nonumber\\
&&{}+O(\theta_0^2).
\eer
\subsection{Propagation dynamics away from {\boldmath $z=z_c$}}
As $\vert z-z_c\vert$ increases, the absolute value of $\vert \gamma\vert$ falls off rapidly and the modulations due to the Hermite polynomials fade away quickly. Using the limiting form $\lim_{\gamma\rightarrow 0} H_m(x/\gamma)=(2x/\gamma)^m$, we can write $E_2(x_2,y_2)$ in terms of incomplete two-variable Hermite polynomials $h_{m,n}(x,y\vert \tau)$, which are defined as \cite{khan, khan4}
\ber
& & \lefteqn{h_{m,n}(x,y\vert \tau) =  m!n!\sum_{j=0}^{\min(m,n)}\frac{\tau^j x^{m-j} y^{n-j}}{j!(m-j)!(n-j)!}}\nonumber \\
 & = &\left\{\begin{array}{ll} m! \tau^m x^{n-m} L_m^{(n-m)} (-xy/\tau), \qquad n>m,\\
 n! \tau^n y^{m-n} L_n^{(m-n)} (-xy/\tau),\qquad m>n.
 \end{array}
 \right.
\eer
Thus $E_2(x_2,y_2)$ reduces to
\ber
& & E_2(x_2,y_2)  = \frac{kw_1w_2i^{m+n+1}}{2(b_1b_2)^{1/2}}\exp\left[-\left(\beta_1 x_2^2+\beta_2 y_2^2\right)\right]\nonumber\\
&&\times
\left\{\begin{array}{ll} m! \tau^m \phi_1^{n-m} L_m^{(n-m)} (-\phi_1\phi_2/\tau), \qquad n>m,\\
 n! \tau^n \phi_2^{m-n} L_n^{(m-n)} (-\phi_1\phi_2/\tau),\qquad m>n.
 \end{array}
 \right.
\eer
where $\phi_j$ are as in Eq. (10) and $\tau=-(w_1^2-w_2^2\epsilon_1\epsilon_2)/2$.
In what follows, we will experimentally demonstrate the validity of our theoretical results.
\section{Experiment}
\begin{figure}[h]
 \begin{center}
 \includegraphics[width=3.5in]{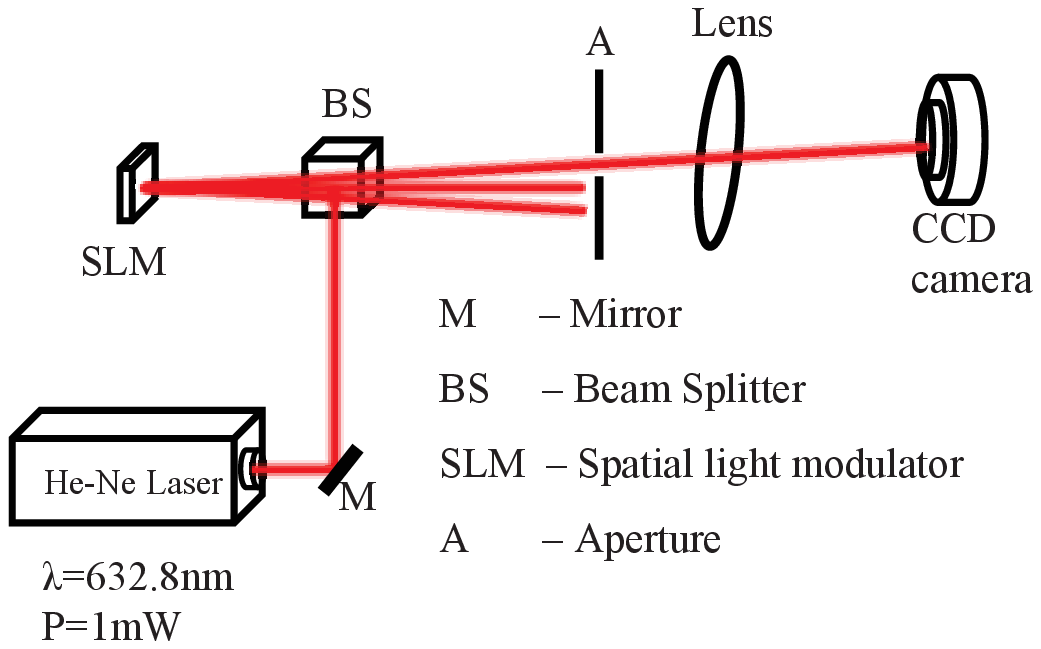}
 \caption{(Colour online) Experimental setup for the determination of the net charge of an arbitrary vortex pair embedded in a Gaussian beam}\label{fig:expt}
 \end{center}
 \end{figure}

The experimental set up  is shown in Fig. \ref{fig:expt}. Suitable phase masks for creating vortex pairs are produced by using computer generated holography (CGH) technique \cite{cgh} and sent to a spatial light modulator (SLM) via a computer.  The SLM is illuminated by an intensity stabilized He-Ne laser (Spectra-Physics, Model 117A) of power $1$ mW and wavelength $632.8$ nm to produce the desired vortex pair. The vortex pair is selected with an aperture (A) and passed through a spherical bi-convex lens of focal length $50$ cm which is tilted by an angle $6^{\circ}$. The tilting of the lens has been done with a rotational stage with least count of $0.1^{\circ}$. The aperture is at a distance $z_1=90$ cm in front of  the SLM. We use the method described in \cite{sirohi}, to find that the Gaussian laser beam hosting the selected vortex pair has a beam waist $0.186$ mm at a virtual point which is at a distance of $z_2=60.8$ cm {\it behind} the SLM. The distance between the lens and the aperture is $z_3=245$ cm. Thus the total distance traveled by the vortex pair from the waist plane to the lens is $z_0=z_1+z_2+z_3=395.8$ cm. The resultant intensity patterns are recorded by a CCD camera (MediaCybernetics, Evolution VF cooled Color Camera) placed at a distance $z$  past the lens.
 \section{Intensity pattern at {\boldmath $z=z_c$} and determination of net topological charge}
 In this section, we determine the net topological charge of the vortex pair from its intensity distribution at $z=z_c$. The predicted value of $z_c$ from Eq. (\ref{zc}) is $57.2$ cm which is close to the experimentally observed value of $56.3$ cm. In the intensity patterns, with reference to Eq. (\ref{vor}), the vortex on the left ($x_1=-x_0$) has a charge $\epsilon_1 m$ and the vortex on the right ($x_1=x_0$) has a charge $\epsilon_2 n$. The corresponding figure is labelled as ($\epsilon_1 m,\,\epsilon_2 n$).
 \subsection{Vortices with topological charges of the same sign}
 \begin{figure}[h]
\begin{center}
\includegraphics[width=3.5in]{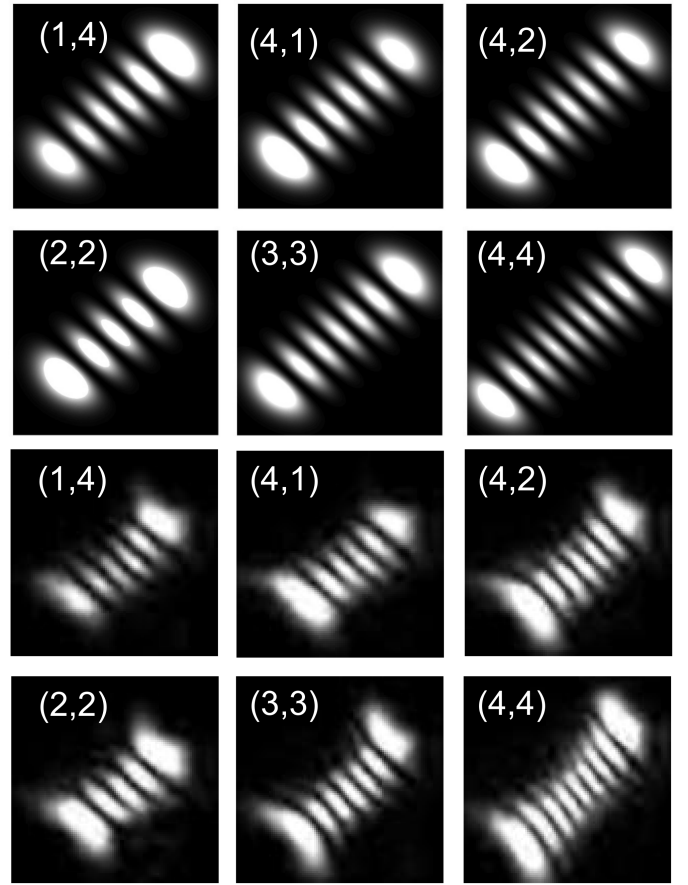}
\caption{  The theoretical (first two rows) and experimental (last two rows) results for the intensity patterns of a vortex pair with topological charges of the same sign, at $z=z_c$ for $x_0=0.1 w_0$.}\label{fig:same1}
\end{center}
\end{figure}

Fig. \ref{fig:same1} shows the theoretical (first two rows) and experimental (last two rows) images for the intensity patterns of a pair of vortices with the same sign ($\epsilon_1=\epsilon_2=1$) but different magnitudes $m$ and $n$ with the separation parameter set at $x_0=0.1 w_0$.

For small separation $x_0$, these patterns can be explained by Eq. (24). Since the first term in (24) is the leading term, one can obtain the net charge $m+n$ by noting that there are $m+n+1$ bright stripes in the intensity distribution. These stripes are parallel to one another and lie along a line that is neither horizontal nor vertical, but tilted in a clockwise direction almost along a diagonal as the dependence on $x$ and $y$ is through a single variable $\theta_- = \delta_1 x_2-\delta_2 y_2$ and $\delta_1\sim \delta_2$. However, interference with the second term will lead to a slightly asymmetric distribution of brightness among the stripes. As is clear from the second term in (24), this asymmetry depends on the difference between the magnitude of charges and the separation between them. Additionally, when the vortices swap their positions as in (4,1) and (1,4), the lower half of the pattern becomes the mirror image of the upper half and vice-versa. For $m=n$ as in (4,4), the two halves have identical intensity patterns. If the charge of each vortex were negative ($\epsilon_1=\epsilon_2=-1$), then $\theta_-$ would be replaced by $\theta_+=\delta_1 x_2+\delta_2 y_2$ and the bright stripes would be tilted in an anti-clockwise fashion (not shown).

\begin{figure}[h]
\begin{center}
\includegraphics[width=3.5in]{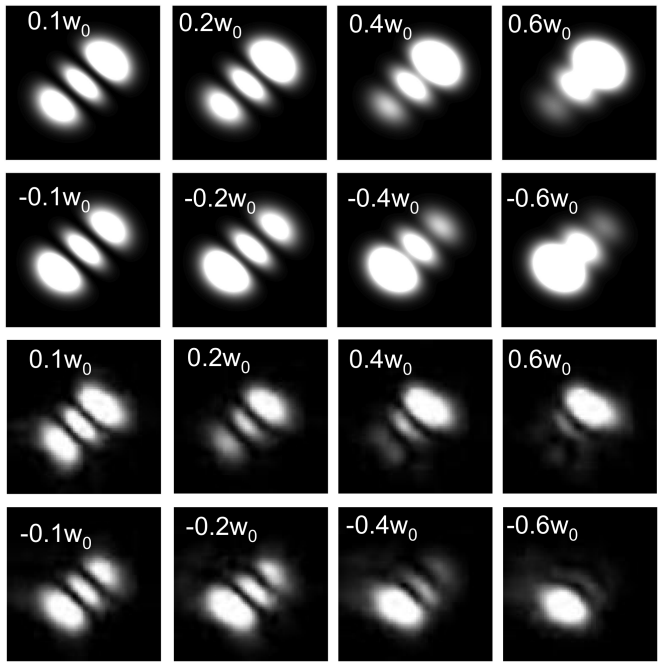}
\caption{  The theoretical (first two rows) and experimental (last two rows) results for the intensity patterns of an off-axis vortex of charge 2, at $z=z_c$ for different values of $x_0$ as labelled in the figures.} \label{fig:off2}
\end{center}
\end{figure}

To investigate the effect of the separation parameter $x_0$, we have also studied the propagation of an off-axis vortex of charge 2 through the tilted lens. The corresponding theoretical  (first two rows) and  experimental (last two rows) results for the intensity patterns at $z=z_c$ are shown in Fig. \ref{fig:off2}. From the images, it is clear that the intensity of one of the outer lobes increases as the vortex moves farther away from the center and the remaining lobes lose their intensity. In the notation of this section, this off-axis vortex can be labeled as (2,0) for $x_0=|x_0|$. Consequently, when $x_0$ becomes negative, the vortex is identically described as (0,2) with $x_0=|x_0|$ and the pattern flips diagonally. The situation is analogous to the case of (4,1) and (1,4) as described in the previous paragraph
\subsection{Vortices with topological charges of opposite signs}

\begin{figure}[h]
\begin{center}
\includegraphics[width=3.5in]{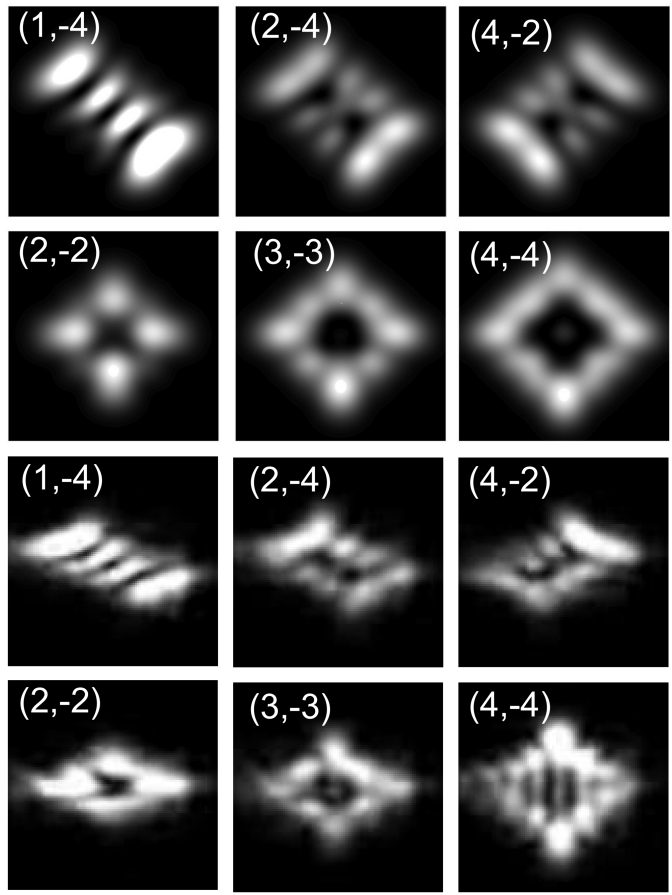}
\caption{ The theoretical (first two rows) and experimental (last two rows) results for the intensity patterns of a vortex pair with topological charges of opposite signs, at $z=z_c$ for $x_0=0.1 w_0$.}\label{fig:opp1}
\end{center}
\end{figure}

Fig. \ref{fig:opp1} shows the theoretical (first two rows) and experimental (last two rows) images corresponding to opposite singularities ($\epsilon_1\epsilon_2=-1$) for separation parameter $x_0=0.1 w_0$ and topological charges as shown in the images.

For small values of $m$ and $n$, these patterns can be explained by expanding the Hermite polynomials in power series. The calculation would be long and tedious. Instead, we make the following empirical observation.
If $m\neq n$, the pattern has a  rectangular `razor-blade' structure which is tilted clockwise (anti-clockwise) if the net charge is positive (negative).  On closer observation, we note that there are $m$ bright spots on two parallel sides and $n$ bright spots on the remaining two parallel sides. Thus, for vortex dipoles ($m=n$) the patten is square with its corners in the east, west, north and south directions, each side having $m=n$ bright spots.

As far as we know, Fig. \ref{fig:opp1} represents {\it the first optical realization} of the 4-variable 2-index 1-parameter Hermite polynomials $H_{m,n}(x,z;y,w|\tau)$ modulated by an elliptical Gaussian beam (see Eqs. 8, 15 and 26).
\begin{figure}[h]
\begin{center}
\includegraphics[width=3.5in]{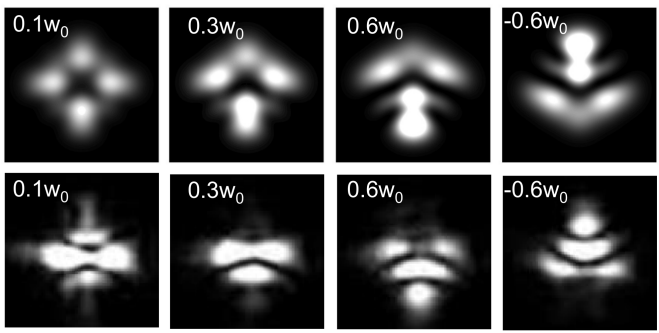}
\caption{  The theoretical (top row) and experimental (bottom row) results for the intensity patterns of a dipole vortex of charge $(2,-2)$, at $z=z_c$ for different values of $x_0$ as labelled in the figures.}
\label{fig:dip2}
\end{center}
\end{figure}

In Fig. \ref{fig:dip2}, we show the evolution of a dipole vortex of charge $(2,-2)$ as a function of separation between the two vortices. For small separation, the intensity distribution is symmetric in both the transverse directions. As the separation is increased, the pattern becomes asymmetric. When the separation parameter $x_0$ becomes negative, the vortex is identically described as $(-2,2)$ with $x_0=|x_0|$ and the pattern flips  vertically.

\begin{figure}[h]
\begin{center}
\includegraphics[width=3.5in]{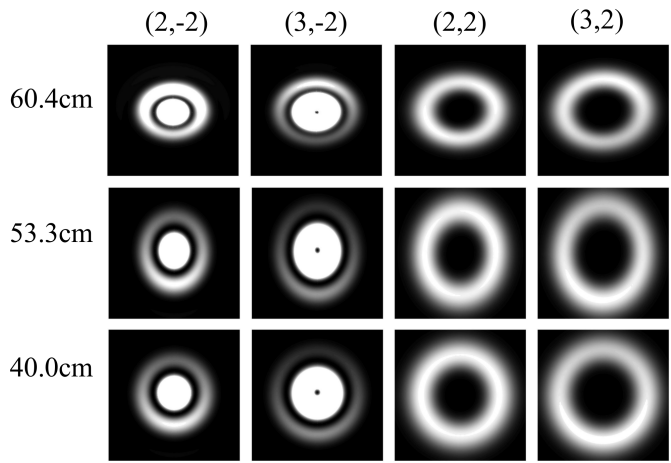}
\caption{  Theoretical intensity patterns of a vortex pair of different charges (as given on the top) at various values of the propagation distance $z$ ( as given on the left).} \label{fig:propt}
\end{center}
\end{figure}

\section{Propagation dynamics away from {\boldmath $z=z_c$}}
\begin{figure}[h]
\begin{center}
\includegraphics[width=3.5in]{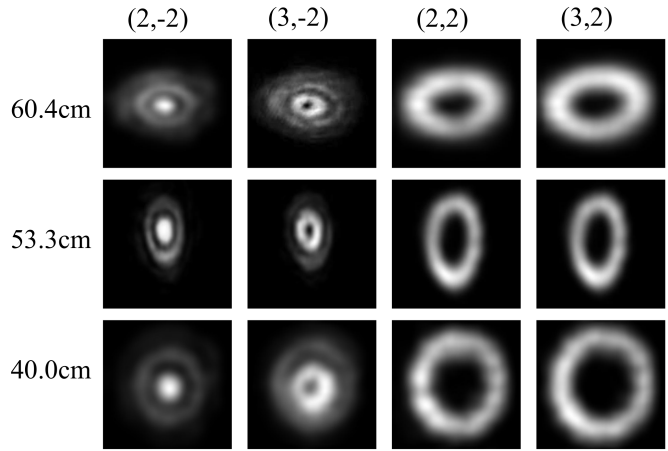}
\caption{  The experimental images corresponding to Fig. \ref{fig:propt}.}
\label{fig:prope}
\end{center}
\end{figure}

As we move away from the point $z=z_c$, the modulations due to the Hermite polynomials disappear quickly. The propagation dynamics is now governed by Eq. (28). The theoretical and corresponding experimental intensity patterns  for various values of $z$ are shown in Figs. \ref{fig:propt} and  \ref{fig:prope} respectively. The intensity patterns are, in general, elliptical. Far away from $z_c$, all patterns become circularly symmetric as $\alpha_1\to \alpha_2$ and $\beta_1\to \beta_2$ \cite{pravin2}.
\pagebreak

\section{Conclusions}
We have studied the propagation of a bi-singular beam with arbitrary topological charges through a tilted lens and used it to find the net topological charge of the beam. This may provide information about the net torque generated by the optical field. This method can be realized easily in the laboratory as it needs just a single tilted lens. Vortices being generic to all the waves, this study may be useful for other systems like acoustic and matter waves.

\end{document}